\documentclass[useAMS,usenatbib,usegraphicx]{mn2e}
\title[Attenuation law through a clumpy ISM]
{Attenuation law of normal disc galaxies with clumpy distributions of stars
  and dust} 
\author[A. K. Inoue]
{Akio K. Inoue\thanks{E-mail:akio.inoue@oamp.fr}\thanks{JSPS Postdoctral
    Fellow for Research Abroad}\\ 
Laboratoire d'Astrophysique de Marseille, Traverse du Siphon, BP8, 13376
Marseille CEDEX 12, France}
\begin{document}

\date{Submitted on 1 December 2004, first revision on 17
  January 2005, second revision on 27 January 2005}

\pagerange{\pageref{firstpage}--\pageref{lastpage}} \pubyear{2005}

\maketitle

\label{firstpage}

\begin{abstract}
We investigate the attenuation law seen through an interstellar medium (ISM)
with clumpy spatial distributions of stars and dust. The clumpiness of the
dust distribution is introduced by a multi-phase ISM model. We solve a set of
radiative transfer equations with multiple anisotropic scatterings through the
clumpy ISM in a 1-D plane-parallel geometry by using the mega-grain
approximation, in which dusty clumps are regarded as very large particles
(i.e.\ mega-grains). The clumpiness of the stellar distribution is introduced
by the youngest stars embedded in the clumps. We assume a smooth spatial
distribution for older stars. The youngest stars are surrounded by denser
dusty gas and suffer stronger attenuation than diffuse older stars (i.e.\
age-selective attenuation). The apparent attenuation law is a composite of the
attenuation laws for the clumpy younger stars and for the diffuse older stars
with a luminosity weight. In general, the stellar population dominating the
luminosity changes from older stars to younger stars as the wavelength
decreases. This makes the attenuation law steep; the composite attenuation
rapidly increases from small attenuation for older stars at a long wavelength
to large attenuation for younger stars at a short wavelength. The resultant
attenuation law of normal disc galaxies is expected to be much steeper than
that of starburst galaxies observed by Calzetti et al. Finally, the Calzetti's
attenuation law is regarded as a special case with a large density in our
framework.
\end{abstract}


\section{Introduction}

Since the first dust grain condensation in the universe, dust grains have
accumulated in the interstellar medium (ISM) of galaxies. This internal dust 
attenuates the stellar light before it escapes from galaxies. Thus, we need a 
correction of the internal dust attenuation in order to obtain the intrinsic
flux and the spectrum of galaxies. This procedure should be done with an
appropriate attenuation\footnote{In this paper, we use the term, 
``attenuation''. However, other terms (e.g., ``effective extinction'', 
``obscuration'') are also used in the literature \citep{cal01}.} law 
including effects of scatterings and the configuration of dust and stars. 
Generally, it is different from the extinction laws observed in the Milky Way
and Magellanic Clouds; an extinction law corresponds to the attenuation law in
a distant uniform screen geometry which is not realistic at all for most
galaxies. However, there are many works in which an extinction law is used to
correct for the internal dust attenuation.

Observationally, only an average attenuation law of ultra-violet (UV) bright 
starburst galaxies has been obtained so far \citep[][ hereafter it is called
the Calzetti law, see also \citealt{lei02} and \citealt{bua02}]{cal94}. 
This is because the UV spectra of galaxies were rare. 
GALEX \citep[GALaxy Evolution eXplorer;][]{mar04} will change the
situation. We will have UV spectra of a lot of galaxies. From the data, we
will obtain attenuation laws of various kinds of galaxies.

Theoretically, to study the attenuation law of galaxies is to solve the
radiative transfer in a galactic scale. There are several groups studying the
galactic radiative transfer \citep[see][ for a review and references
therein]{cal01}. We will present a brief summary for the works about the
attenuation law below. Because of the importance of scatterings
\citep[e.g.,][]{bru88}, we restrict ourselves to only works which include 
scatterings.

The attenuation law through a smooth dust and stellar distribution have been 
examined since the early phase of this topic \citep{bru88,dib95,fer99,bae01b}. 
In general, the attenuation amount is smaller than that by a distant uniform
screen because of the geometric and scattering effects.
For example, let us compare a plane-parallel slab where dust and stars are
well mixed with the foreground uniform screen which has the same amount of dust
as the slab. Obviously, the optical depth from an outside observer to stars
close to the surface of the slab is smaller than that of the foreground screen.
This is a geometric effect. 
The scattering effect is that some scattered photons coming into an observer's
line of sight compensate a part of the radiation absorbed and scattered out
from the line of sight. Therefore, the geometric and scattering effects make
the medium less opaque.

The clumpiness of the medium (i.e. dust distribution) is also important as
clearly shown by \cite{nat84}. This also makes the medium more transparent
than a uniform screen geometry because there are photons escaping from the
medium through a less opaque path \citep{gor97,var99,gor00,wit00,bia00,pie04}. 
Since such effects are more efficient for shorter wavelengths, the resultant
attenuation law usually becomes grayer than the extinction law (distant
uniform screen). For example, 
\cite{gor97} showed that the attenuation law through the clumpy medium with
the Small Magellanic Cloud (SMC) type dust (i.e. no prominent feature at 2175
\AA) can be very similar to the Calzetti law which is much grayer than the
assumed SMC extinction law.

We can also consider the clumpiness of the stellar distribution. As introduced
by \cite{sil98}, young stars may localize in their birth clouds
(i.e. molecular clouds). Because of a larger density of the surrounding gas,
the radiation from young stars suffers stronger attenuation than that from
stars distributed diffusely outside the birth clouds. With this hypothesis
(hereafter it is called the age-selective attenuation), \cite{gra00} well
reproduced the Calzetti law with the Milky Way (MW) type dust (i.e. with the
2175 \AA\ bump). For starburst galaxies, the radiation field is
dominated by the young stellar radiation which suffers a featureless and
grayer attenuation law expected through a dense medium. \cite{gra00} also
obtained a different attenuation law for normal galaxies where the diffuse
component plays a role. \cite{tuf04} showed that the global attenuation law is
a composite of attenuation laws of various stellar components with weights of
their fluxes.

However, there has been no study of the attenuation law taking into account
the clumpiness for both distributions of dust and stars, except for
\cite{bia00}. Since their analysis is monochromatic, it is not the
attenuation law in a strict sense. In real galaxies, the ISM is very clumpy,
and the localization of young stars is often observed. Therefore, we should
treat the clumpiness for both of stars and dust to develop the investigation
of the attenuation law. The current paper executes it for the first time.

This paper also discusses the relation between the clumpiness and the physics
of the medium. In the radiative transfer through the clumpy two-phase medium,
there are two essential parameters: the density contrast between clumps and
the inter-clump medium, and the volume filling fraction of clumps
\citep[e.g.,][]{wit96}. We translate these two parameters into more
fundamental quantities, the ISM thermal pressure and the mean density, based
on the multi-phase ISM model \citep[e.g.,][]{fie69}; we try to relate the
attenuation law with the ISM physics. Recently, such an attempt was started by
\cite{fis03,fis04}. Since their analysis was only a distant foreground screen
(not uniform but turbulent) case, the geometry considered in this paper is
more suitable for extended sources like normal disc galaxies.

This paper is organized as follows: in section 2, we introduce a
physical model of the clumpiness as an application of the two-phase ISM
model. Then, the radiative transfer and the mega-grain approximation
\citep{var99} are described in section 3. The main result of this paper, a
steep attenuation law, is presented in section 4. The robustness of the
result, the Calzetti law, and a simple prescription of the dust attenuation
proposed previously are discussed in section 5. The final section is devoted
to the conclusion of this paper.

\section{Clumpiness of the ISM}

The ISM in galaxies is very clumpy, and the clumpiness is related with the
physics of the ISM. As an application of the multi-phase ISM model
\citep[e.g.,][]{fie69,mck77}, a physical model of the clumpiness is introduced
in this section.

\begin{figure}
  \includegraphics[width=\linewidth]{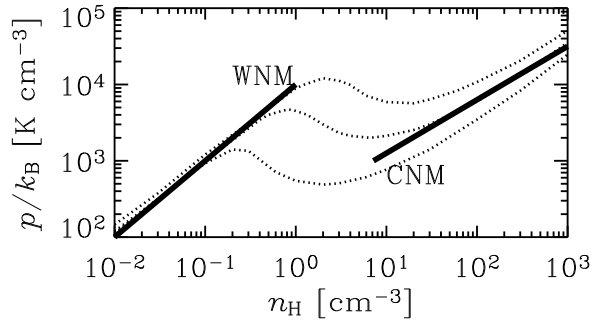}
  \caption{Phase diagram: thermal pressure--hydrogen number density. The 
    dotted curves indicate the thermal equilibrium points in the interstellar
    medium of the Milky Way (top: Galactocentric radius of 3 kpc, middle: 8.5
    kpc, and bottom: 15 kpc) reproduced from fig.7 of Wolfire et al.~(2003). 
    The thick solid lines are the approximate relations of two thermally
    stable phases; the warm neutral medium (WNM, left) and the cold neutral
    medium (CNM, right). Their analytical forms are given by equations (1) and
    (2).}
\end{figure}

Assuming thermal energy and chemical equilibria in the ISM with a 
temperature less than $10^4$ K, we find two thermally stable phases in the
pressure-density phase diagram \citep[e.g.,][]{wol95,koy00}: one is the warm
neutral medium (WNM) and the other is the cold neutral medium (CNM). Figure 1
is an example of the phase diagram. The dotted curves indicate 
the equilibrium points in environments of the Milky Way's ISM calculated by
\cite{wol03}. These curves depend on the heating and cooling functions
considered; they depend on the metallicity, the dust amount and properties,
the intensity of the radiation field, the flux of cosmic rays, and so on. 
However, differences are apparent mainly in the
local maximum and minimum pressures; locations of thermally stable phases 
(especially for the WNM) are rather robust. Thus, we adopt the following
approximate relations between the thermal pressure and the hydrogen number
densities in the two phases: 
\begin{equation}
  \frac{p/k_{\rm B}}{\rm 10^4\,K\,cm^{-3}} = 
  \frac{n_{\rm H,wnm}}{\rm 1\,cm^{-3}}\hspace{0.5cm}{\rm (WNM)},
\end{equation}
and 
\begin{equation}
  \frac{p/k_{\rm B}}{\rm 10^{4.5}\,K\,cm^{-3}} = 
  \left(\frac{n_{\rm H,cnm}}{\rm 10^3\,cm^{-3}}\right)^{0.7}
  \hspace{0.5cm}{\rm (CNM)}.
\end{equation}
They are shown in figure 1 as thick solid lines.

We also assume a thermal pressure equilibrium among phases. Although the total
ISM pressure is dominated by the non-thermal (turbulent motions, magnetic
fields, and cosmic rays) pressure \citep[e.g.,][]{bou90}, thermal pressures of
warm and cold neutral gases and hot gas are similar observationally, 
$p/k_{\rm B}=10^{3-4}$ K cm$^{-3}$ \citep[e.g.,][]{mye78}. 
\cite{wol03} proposed a scenario where the thermal energy and pressure
equilibria, and then, the two-phase neutral medium should be established in
the Milky Way's ISM even in a turbulent condition.
Such equilibria are probably valid for the ISM in other normal disc
galaxies, at least as a long term averaged property.

In the radiative transfer through a two-phase medium discussed here, we regard
the WNM and the CNM as the inter-clump medium and clumps, respectively. 
Here, the total filling fraction of the WNM and the CNM
is assumed to be unity for simplicity although it is about 0.5
observationally \citep{hei03}. In fact, the rest $\sim$50\% of the volume is
filled with the hot ($\sim 10^6$ K) gas produced by supernovae. This hot gas 
determines the thermal pressure of the WNM and the CNM \citep{mck77}.

Once an equilibrium thermal pressure in the medium is given, we have the
densities of the CNM and the WNM based on equations (1) and (2). In other
words, we have the density contrast between clumps and the inter-clump
medium. Then, we assume a mean density of the hydrogen nucleus which gives the
volume filling fraction of clumps; 
$f_{\rm cl}=(n_{\rm H}-n_{\rm H,wnm})/(n_{\rm H,cnm}-n_{\rm H,wnm})$.

Here, we deal with normal disc galaxies. An equilibrium thermal pressure of 
$p/k_{\rm B}=10^{3.5}$ K cm$^{-3}$  \citep[][ for the solar vicinity]{wol03} 
and a mean hydrogen number density of $n_{\rm H}=1$ cm$^{-3}$ 
\citep{spi78} are assumed. These values are listed in Table 1 where other
quantities assumed later are also summarized. The corresponding density
contrast and clump filling fraction are 118 and $1.85\times10^{-2}$,
respectively. This density contrast is similar to that considered in
\cite{wit00}, whereas our filling fraction is about an order of magnitude
smaller than theirs. That is, our mean density of the medium is much less than
that in \cite{wit00}.

Finally, clumps are assumed to be self-gravitating; the clump size, which is
required to calculate the optical depth of each clump later, is given by the
equilibrium pressure and the CNM density through the Jeans length. This is an
assumption motivated by the following discussion where we regard
clumps as birth clouds of young stars. Indeed, the molecular clouds should be
in the CNM. For the equilibrium pressure and mean density assumed above, the
Jeans length for a sphere of the CNM is  
$\sqrt{15 p/4\pi G}/\rho_{\rm cl} = 10.4$ pc ($=r_{\rm cl}$), 
where $p$ is the thermal pressure, $G$ is the gravitational constant, and the
density inside clumps $\rho_{\rm cl} = \mu m_{\rm p} n_{\rm H,cnm}$ with the
mean atomic weight $\mu=1.4$ and the proton mass $m_{\rm p}$. This size is
similar to (but slightly larger than) a typical size of the molecular clouds
\citep[e.g.,][]{lar81}.

\begin{table}
 \centering
 \begin{minipage}{\linewidth}
  \caption{Physical quantities determining the attenuation law.}
  \begin{tabular}{ccc}
   \hline
   Name & Notation & Standard value\\
   \hline
   equilibrium thermal pressure & $p/k_{\rm B}$ & $10^{3.5}$ K cm$^{-3}$\\
   mean hydrogen density & $n_{\rm H}$ & 1 cm$^{-3}$\\
   dust-to-gas mass ratio & $\cal D$ & $10^{-2}$\\
   half height of the gas+dust disc & $h_{\rm d}$ & 150 pc\\
   layering parameter of old stars & $\xi$ & 0.5\\
   young stellar age criterion & $t_{\rm y}$ & $10^7$ yr\\
   \hline
   \multicolumn{2}{l}{homogeneous visual optical depth (face-on)} & 0.54\\
   \hline
  \end{tabular}
 \end{minipage}
\end{table}

\section{Radiative Transfer}

\subsection{Dust properties}

In this paper, we use an empirical set of dust optical properties taken from
\cite{wit00}. This data set consists of the dust opacity 
($k_{\rm d,\lambda}$), the scattering albedo ($\omega_{\rm d,\lambda}$), and 
the asymmetry parameter ($g_{\rm d,\lambda}$) for the MW type and the SMC type
dust. The number of wavelengths is 25 from 0.1 \micron\ to 3 \micron.  Since
there is only the wavelength dependence for $k_{\rm d,\lambda}$ in the data
set, we adopt the absolute value of the opacity at visual band as 
$k_{{\rm d},V}=2.5\times10^4$ cm$^2$ g$^{-1}$ 
\citep[extinction cross section per unit dust mass; e.g.,][]{dra03} 
for both types of dust although the actual values for the MW dust and
the SMC dust are slightly different.
Effects of the size distribution and the composition are already included in
the data. For calculations of the scattering angle, we assume the
Henyey-Greenstein phase function \citep{hen41} for simplicity although
\cite{dra03} proposed a better function. Finally, we assume the dust-to-gas
mass ratio in the ISM of the Milky Way ${\cal D} = 10^{-2}$
\citep[e.g.,][]{spi78}, as a standard value.

\subsection{Mega-grain approximation}

To treat clumpiness in the radiative transfer problem, we have to consider a 
spatial 3-D geometry. However, we can do in a spatial 1-D geometry with the
mega-grain approximation introduced by \cite{neu91,hob93} and further
developed by \cite{var99}. In this approximation, we regard a dusty clump as a
huge particle called a mega-grain which produces absorption
and scattering effects like a normal dust grain. Practically, we replace usual
optical properties of normal dust grains (extinction coefficient, albedo, and
asymmetry parameter) with effective ones. We follow the formulation given by
\cite{var99}, in which an extensive comparison between the approximation and 
3-D Monte-Carlo radiative transfer calculations has been made in some spherical
geometries and the validity of the approximation has been shown clearly. We
give a summary of equations used in this paper below. We will omit to write
the wavelength dependence explicitly.

We assume that all clumps are an identical sphere with the radius $r_{\rm cl}$
and the gas density $\rho_{\rm cl}$ given in section 2 and these clumps are
floating in the inter-clump medium with the gas density 
$\rho_{\rm icm} = \mu m_{\rm p} n_{\rm H,wnm}$. The optical depth (radius) of
a clump relative to the inter-clump medium is 
\begin{equation}
  \tau_{\rm cl} = (\rho_{\rm cl} - \rho_{\rm icm}) 
  k_{\rm d} {\cal D} r_{\rm cl}\,.
\end{equation}
This optical depth vanishes in the case of no density contrast. Since the
interaction (absorption and scattering) probability against the parallel light
by a sphere with an optical depth (radius) $\tau$ is exactly 
\begin{equation}
  P_{\rm int}(\tau) = 1 - \frac{1}{2\tau^2} 
  + \left(\frac{1}{\tau} + \frac{1}{2\tau^2}\right)e^{-2\tau}\,, 
\end{equation}
the extinction coefficient per unit length of the medium by clumps
(i.e. mega-grains) is  
\begin{equation}
  \kappa_{\rm mg} = n_{\rm cl} \pi r_{\rm cl}^2 P_{\rm int}(\tau_{\rm cl})
  = \frac{3f_{\rm cl}}{4r_{\rm cl}}P_{\rm int}(\tau_{\rm cl})\,,
\end{equation}
where the clump number density $n_{\rm cl}$ was replaced with 
$3f_{\rm cl}/4\pi r_{\rm cl}^3$. We did not consider any overlap of clumps.
Then, the effective extinction coefficient per unit length
in the two-phase medium is defined as 
\begin{equation}
  \kappa_{\rm eff} = \kappa_{\rm mg} + \kappa_{\rm icm}\,,
\end{equation}
where $\kappa_{\rm icm} = k_{\rm d}{\cal D}\rho_{\rm icm}$ is the extinction
coefficient per unit length of the inter-clump medium.

The effective albedo $\omega_{\rm eff}$ and the effective asymmetry parameter
$g_{\rm eff}$ are also defined as 
\begin{equation}
  \omega_{\rm eff} = \frac{\omega_{\rm cl}\kappa_{\rm mg} 
    + \omega_{\rm d}\kappa_{\rm icm}}{\kappa_{\rm eff}}\,,
\end{equation}
and 
\begin{equation}
  g_{\rm eff} = \frac{g_{\rm cl}\kappa_{\rm mg} 
    + g_{\rm d}\kappa_{\rm icm}}{\kappa_{\rm eff}}\,,
\end{equation}
respectively, where $\omega_{\rm cl}$ and $g_{\rm cl}$ are the albedo and
asymmetry parameter of a clump, respectively. \cite{var99} proposed an
expression of $\omega_{\rm cl}$ as 
\begin{equation}
  \omega_{\rm cl} = \omega_{\rm d} 
  P_{\rm esc}(\tau_{\rm cl},\omega_{\rm d})\,, 
\end{equation}
where 
\begin{equation}
  P_{\rm esc}(\tau,\omega) = 
  \frac{(3/4\tau)P_{\rm int}(\tau)}
  {1-\omega[1-(3/4\tau)P_{\rm int}(\tau)]}\,.
\end{equation}
This is the photon escape probability from a sphere in which isotropic sources
and dust whose scattering is isotropic distribute uniformly. In the
derivation of equation (10), the uniform distribution of the scattered photons
is also assumed \citep[see Appendix C of ][]{var99}. Although equation (9) is
an approximation, the agreement with results obtained by the Monte Carlo
simulation is good \citep{var99}. For the clump asymmetry parameter, we use
equation (43) of \cite{var99} which is an analytical fitting formula of  
their Monte Carlo simulation results.

In the case with a large filling fraction of clumps $f_{\rm cl}$, the overlap
of clumps happens. \cite{var99} proposed the replacement of $r_{\rm cl}$
with $r_{\rm cl} (1-f_{\rm cl})^\gamma$ to take into account this overlap
effect. \cite{var99} also noted that $\gamma=1$ is suitable for the case of
uniformly distributed sources which is our case. Although it is included in
the computation, this effect is almost negligible for the standard case
discussed later because $f_{\rm cl} \ll 1$.

\subsection{Transfer equations}

Suppose a 1-D plane-parallel geometry along $z$-axis. The mid-plane is set to
be $z=0$ where we put a mirror boundary. Above the mid-plane, we put the
gas+dust disc with a constant mean density up to a height of $h_{\rm d}$. 
Here, we assume $h_{\rm d}=150$ pc which is the observed height of the cold
neutral and molecular gas in the Milky Way \citep[e.g.,][]{bin98}.
Although its mean density is constant, it has a clumpy structure described 
in section 2, and its extinction coefficient is the effective one 
$\kappa_{\rm eff}$ given by the mega-grain approximation. For simplicity, we
assume that there is no dust outside the gas+dust disc. Then, the optical
depth coordinate is defined as $d\tau=-\kappa_{\rm eff} dz$ with $\tau=0$ at 
$z=h_{\rm d}$. Since $\kappa_{\rm eff}$ is constant throughout the gas+dust
disc, we have $\tau=\kappa_{\rm eff}h_{\rm d}$ at $z=0$. Note that the
effective albedo $\omega_{\rm eff}$ and the effective asymmetry parameter 
$g_{\rm eff}$ are also constant throughout the disc.

A static radiative transfer equation in this geometry is  
\begin{equation}
  \mu \frac{dI(\tau,\mu)}{d\tau} = S(\tau,\mu) - I(\tau,\mu)\,,
\end{equation}
where $I$ is the specific intensity along a ray and $\mu$ is the cosine of the
angle between the ray and the $z$-axis. The source function $S$ is given by 
\begin{equation}
  S(\tau,\mu) = \eta_*(\tau) / \kappa_{\rm eff}
  + \omega_{\rm eff} \int_{-1}^1 I(\tau,\mu') \Phi(g_{\rm eff},\mu,\mu') d\mu'\,,
\end{equation}
where $\eta_*$ is the stellar emissivity (isotropic) and $\Phi$ is the
scattering phase function. Note that $\omega_{\rm eff}$ and $g_{\rm eff}$
appear in the second term of the right hand side (i.e. scattering term). 
The boundary conditions are 
\begin{equation}
  I(\tau=\kappa_{\rm eff}h_{\rm d},\mu) = I(\tau=\kappa_{\rm eff}h_{\rm d},-\mu)\,,
\end{equation}
at $z=0$ (mirror boundary) and 
\begin{equation}
  I(\tau=0,\mu<0) = -\frac{I_*}{\mu}\,,
\end{equation}
at $z=h_{\rm d}$, where
\begin{equation}
  I_* \equiv \int_{h_{\rm d}}^\infty \eta_*(z) dz\,.
\end{equation}

\subsection{Emissivity distribution --- Stellar clumpiness}

We consider two kinds of the emissivity source: one is embedded in clumps and
the other distributes diffusely and smoothly. Then, we consider that the
embedded source is the youngest stars (the age criterion is defined later) and
other older stars distribute diffusely. This is motivated by the fact that
stars are formed in molecular clouds (i.e. clumps) and diffuse as ageing. This
makes the clumpiness of the emissivity. Here we normalize the intrinsic
(i.e. in the no dust case) emissivity of each component as  
\begin{equation}
\int_{-\infty}^\infty\eta_*(z)dz=1\,.
\end{equation}

\subsubsection{Clumpy stellar emissivity}

When clumps distribute uniformly in the gas+dust disc (we did not consider
any spatial correlation of clumps in section 2), a mean density of stars
embedded in clumps is constant throughout the gas+dust disc. On the other
hand, there is no embedded star outside the disc because there is no
clump. Thus, the intrinsic clumpy stellar emissivity is $1/2h_{\rm d}$ for  
$|z| \leq h_{\rm d}$ and zero for $|z| > h_{\rm d}$ when it is normalized as 
equation (16). A part of the radiation energy from the embedded stars is
locally absorbed by dust in a clump. If stars distribute in the clump
uniformly and the scattering by dust is isotropic, we can adopt the escape
probability of equation (10). Therefore, the clumpy stellar emissivity input
into the transfer equation is 
\begin{equation}
  \eta_*(z) = \cases{
    P_{\rm esc}(\tau_{\rm cl},\omega_{\rm cl})/2h_{\rm d}
    & (for $|z| \leq h_{\rm d}$) \cr
    0 & (for $|z| > h_{\rm d}$) \cr}\,.
\end{equation}
In this case, we have $I_*=0$.

\subsubsection{Diffuse stellar emissivity}

The diffuse stellar disc is assumed to have an exponential structure
along $z$-axis with a scale height of $h_{\rm d}/\xi$, where $\xi$ is the
layering parameter, i.e. the ratio of the heights of the dusty disc to the
stellar disc. Here we set $\xi=0.5$ which means that the stellar scale height
is twice times larger than the height of the gas+dust disc
\citep[e.g.,][]{bin98}. From the normalization of equation (16), we have 
\begin{equation}
  \eta_*(z) = \frac{\xi e^{-\xi |z|/h_{\rm d}}}{2h_{\rm d}}\,.
\end{equation}
In this case, $I_*=e^{-\xi}/2$.

\subsection{Computational scheme}

The radiative transfer equation described above was solved by the
$\Lambda$-iteration scheme with a difference formula of the formal solution 
\citep[e.g.,][]{mih99}. The iteration was continued until the maximum
fractional difference between previous and present values of $S$ becomes less
than $10^{-10}$. To accelerate the convergence, we adopted the Ng's algorithm
\citep{ng74,ols86}. Thanks to a modest optical depth of the problem discussed
here, the number of iterations was always less than 10--20 times. 

As a code test, we compared the inclination dependences of the attenuation
amount for several optical bands in a smooth plain-parallel geometry
calculated by the radiative transfer code of this paper with that calculated
by a spherical harmonics method presented in table 2 of \cite{bae01a}. For the
comparison, we adjusted the code to the set up of \cite{bae01a}. The agreement
was excellent: differences were within a few \% in magnitude.

\subsection{Transmission rate}

We first define the transmission rate for each emissivity source.
The intensity observed by a distant observer (in a direction $\mu>0$) is 
\begin{equation}
  I_{\rm obs}(\mu>0) = I(\tau=0,\mu>0) + \frac{I_*}{\mu}\,.
\end{equation}
On the other hand, the intrinsic (no dust case) intensity to the observer is 
\begin{equation}
  I_{\rm int}(\mu>0) = \frac{1}{\mu}\,, 
\end{equation}
because of the normalization of equation (16). Thus, the transmission rate is 
\begin{equation}
  T(\mu>0) \equiv \frac{I_{\rm obs}(\mu>0)}{I_{\rm int}(\mu>0)}
  = \mu I(\tau=0,\mu>0) + I_*\,.
\end{equation}
When we consider two components of the emissivity source (clumpy and diffuse),
the total transmission rate becomes a composite of transmission rates for
them as 
\begin{equation}
 T = f T_1 + (1-f) T_0\,,
\end{equation}
where $f$ is the fraction of the emissivity embedded in clumps, and 
\begin{description}
\item $T_1$: The transmission rate for the clumpy source.
\item $T_0$: The transmission rate for the diffuse source.
\end{description}
Hereafter, these abbreviations are used.

Since we consider that the embedded source is the youngest stars and the
diffuse source is older stars, $T_1$ and $T_0$ are transmission rates for the
radiation from younger stars inside clumps and for the radiation from older
stars outside clumps, respectively.
The emissivity fraction $f$ is now the luminosity fraction of the younger
stars. We denote it as $f_{\rm y}$. It depends on the star formation history
in addition to the age criterion of young stars. If the star formation rate at
a galactic age $t$ is denoted by $\Psi(t)$, the luminosity fraction $f_{\rm y}$
as a function of the age $t$ is 
\begin{equation}
  f_{\rm y}(t) = \frac{\int_0^{\min[t_{\rm y},t]} \Psi(t-t')L(t') dt'}
  {\int_0^{t} \Psi(t-t')L(t') dt'}\,,
\end{equation}
where $t_{\rm y}$ is the criterion of the young stellar population, and 
$L(t)$ is the luminosity density of the simple stellar population at the
age $t$.

\begin{figure}
  \includegraphics[width=0.9\linewidth]{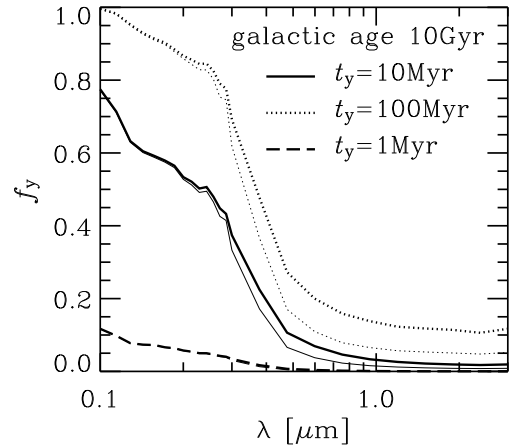}
  \caption{Luminosity fraction of young stellar population as a function of
    the wavelength at the galactic age of 10 Gyr. The solid, dashed, and
    dotted curves are the cases with the age criterion for young stars of 10
    Myr, 1 Myr, and 100 Myr, respectively. A constant star formation history
    is assumed for thick curves, and an exponentially decaying star formation
    history with 5 Gyr time-scale is assumed for thin curves. The difference
    between the two dashed curves are too small to see it.}
\end{figure}

Figure~2 shows the wavelength dependence of $f_{\rm y}$ for several values of 
$t_{\rm y}$ in the galactic age of 10 Gyr case. Although we assumed two star
formation histories, a constant star formation rate and an exponentially
decaying star formation with 5 Gyr time-scale, the differences are very
small. Hereafter, we discuss only the constant star formation case.
On the other hand, the choice of $t_{\rm y}$ is more important.

The age criterion, $t_{\rm y}$ is physically related with the life-time of the
birth clouds (i.e. clumps) and/or the time-scale of stars moving out from the
birth clouds. The life-time of the molecular clouds is estimated to be about
10 Myr \citep{bli80}. The moving off time-scale is also about 10 Myr for the
size of clumps considered here since the observed velocity dispersion of young
stars is $\sim 1$ km s$^{-1}$ \citep[e.g.,][]{deb99}. Moreover, 
\cite{cal94} found larger attenuation against the hydrogen recombination
lines than that against the UV stellar continuum. This implies that ionizing
stars (i.e. O-type stars) are more deeply embedded in dusty clouds than
later-type stars. The main-sequence life-time of the latest O-type stars is
also about 10 Myr. Therefore, we choose 10 Myr as a typical value for 
$t_{\rm y}$ \citep{sil98,cha00}.

\section{Result: a steep attenuation law}

\begin{figure*}
  \includegraphics[width=0.9\linewidth]{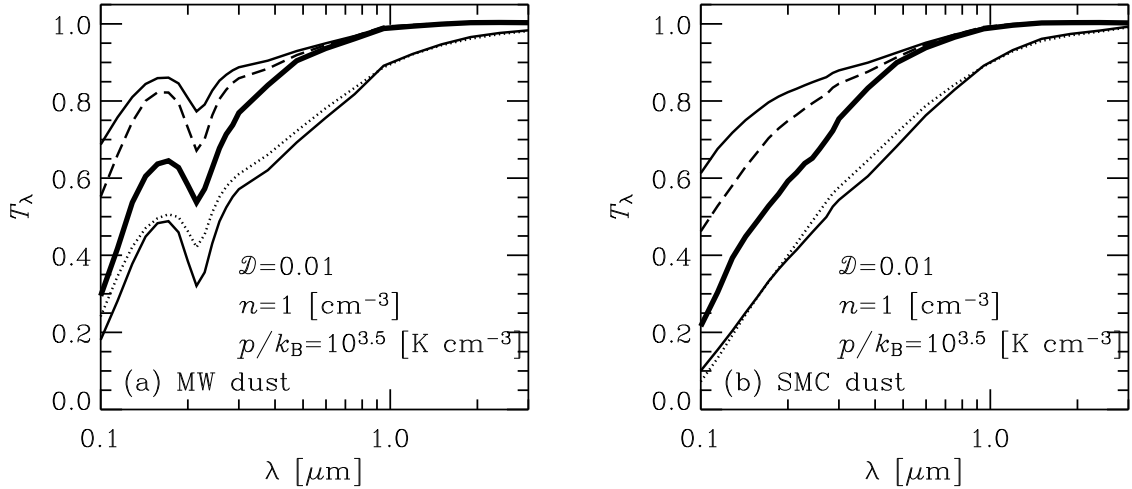}
  \caption{Face-on transmission rates for (a) the Milky Way type
    dust and (b) the Small Magellanic Cloud type dust. In each panel, the
    upper thin solid curve is the transmission rate for the source outside
    clumps ($T_0$), the lower thin solid curve is that for the source inside
    clumps ($T_1$), and the thick solid curve is a composite of the two cases
    with the luminosity weight $f_{\rm y}$ displayed as the solid curve in
    figure~2 (the age criterion of young stars is 10 Myr). The dotted and
    dashed curves are the uniform screen and the uniform disc (no clumpiness
    for stars and dust) cases, respectively. The height of the dusty disc is
    150 pc and the layering parameter of the diffuse stellar disc is
    0.5. Other assumed quantities are displayed in the panels.}
\end{figure*}

\begin{figure*}
  \includegraphics[width=0.9\linewidth]{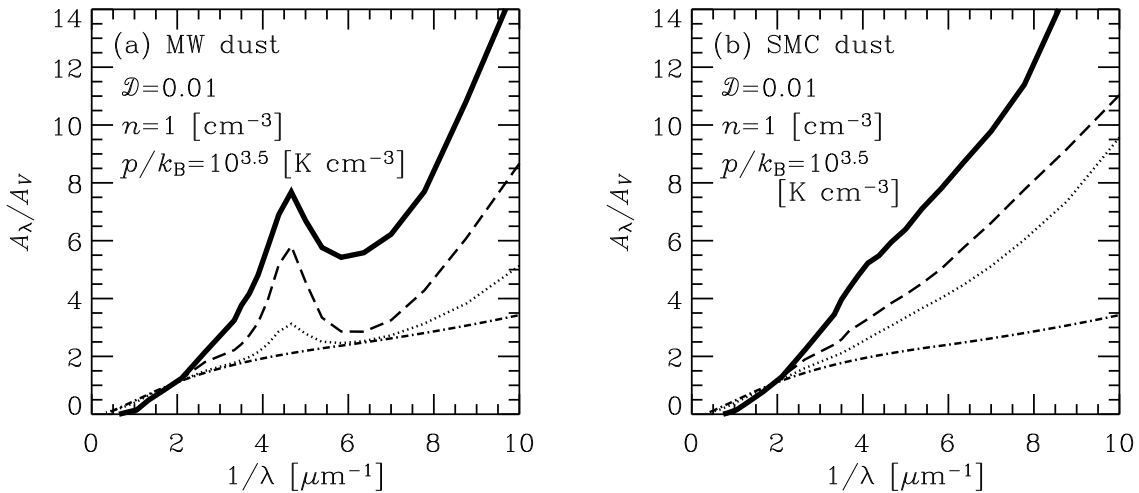}
  \caption{Face-on attenuation laws (normalized at visual band) for (a) the
    Milky Way type dust and (b) the Small Magellanic Cloud type dust. The
    solid curve is the typical case for normal disc galaxies denoted as the
    thick solid curve in figure~3. The dotted and dashed curves are the
    uniform screen and the uniform disc (no clumpiness for stars and dust)
    cases, respectively. The dot-dashed curve is the Calzetti law
    \citep{cal01,lei02}. The height of the dusty disc is 150 pc and the layering
    parameter of the diffuse stellar disc is 0.5. The criterion of the young
    stellar age is 10 Myr. Other assumed quantities are displayed in the
    panels.}
\end{figure*}

We first show a typical attenuation law of normal disc galaxies with the
standard values of physical quantities summarized in table 1.
For comparisons with the literature, we note the equivalent face-on visual
optical depth in the homogeneous case: 
\begin{equation}
  \tau^{\rm hom}_V = 0.54 
  \left(\frac{n_{\rm H}}{1\,{\rm cm^{-3}}}\right)
  \left(\frac{h_{\rm d}}{150\,{\rm pc}}\right) 
  \left(\frac{\cal D}{10^{-2}}\right)\,, 
\end{equation}
for the visual extinction cross section per unit dust mass 
$k_{{\rm d},V}=2.5\times10^4$ cm$^2$ g$^{-1}$.

Figure~3 shows the transmission rates for the face-on ($\mu=1$) case: the
panel (a) is the case assumed the MW type dust (with the 2175 \AA\ bump)
and the panel (b) is the case assumed the SMC type dust (without the bump).
The solid curves are the cases of the clumpy dust distribution: in each panel, 
the upper thin curve is the transmission rate $T_0$ and the lower thin curve
is the transmission rate $T_1$. The thick solid curve in each panel is the
typical case for normal disc galaxies, i.e. the composite transmission rate by
equation (22) with $f_{\rm y}$ displayed as the solid curve in figure~2
($t_{\rm y}=10$ Myr). For a comparison, the uniform screen case and the
uniform gas+dust disc case (i.e. no clumpiness for stars and dust) are shown
by the dotted and dashed curves, respectively.

Generally the ISM clumpiness makes the medium less opaque. Indeed, the
transmission rates increase in order of the screen (dotted), the uniform
disc (dashed), and the clumpy disc (upper thin solid, $T_0$). However, we can
have heavier attenuation if stars are embedded in clumps (lower thin solid,
$T_1$) \citep[see also][]{bia00}.

The composite transmission rate, $T_\lambda$ follows
$T_{0,\lambda}$ at long wavelengths where old stars outside clumps dominate
the luminosity (i.e., $f_{\rm y,\lambda}\simeq0$), whereas it approaches 
$T_{1,\lambda}$ at short wavelengths where young stars inside clumps dominate
the luminosity (i.e., $f_{\rm y,\lambda}\simeq1$). 
This makes the attenuation law steep. As shown in figure~4, 
the expected typical attenuation law of normal disc galaxies (thick
solid curve) is much steeper than the Calzetti law\footnote{We extrapolated 
the attenuation law presented by \cite{cal01}, based on \cite{lei02} for 
$\lambda < 0.18 \micron$.} (dot-dashed curve), the extinction laws (i.e.,
uniform screen case, dotted curve) and the attenuation law for the uniform
stellar and dust distribution (dashed curve).
This is the main conclusion of this paper.

Before moving to the examination of the parameter dependences, we note that 
the transmission rate slightly larger than unity for a very small opacity case
(e.g., longer wavelengths). For the case displayed in figure~3, the
transmission rate is 1.004 at 2 \micron, that is, 0.4\% brighter than the
intrinsic luminosity. This is not artefact but real due to the effect of
scatterings \citep[e.g.,][ see also section 5.1.7]{bae01b}.

\section{Discussion}

\subsection{Robustness of the result}

A steep attenuation law obtained in the previous section is mainly caused by
the age-selective attenuation; a composite of two different attenuations,
namely, smaller attenuation for diffuse stars and larger 
attenuation for young stars embedded in clumps. Here, we examine which
parameters affect on the conclusion by changing values of them from the
standard case (table 1) one by one. In figures of this section, we will see
only one case of the MW type dust or the SMC type dust because the two dust
cases give qualitatively very similar results.

\subsubsection{Young stellar age criterion}

\begin{figure}
  \includegraphics[width=0.9\linewidth]{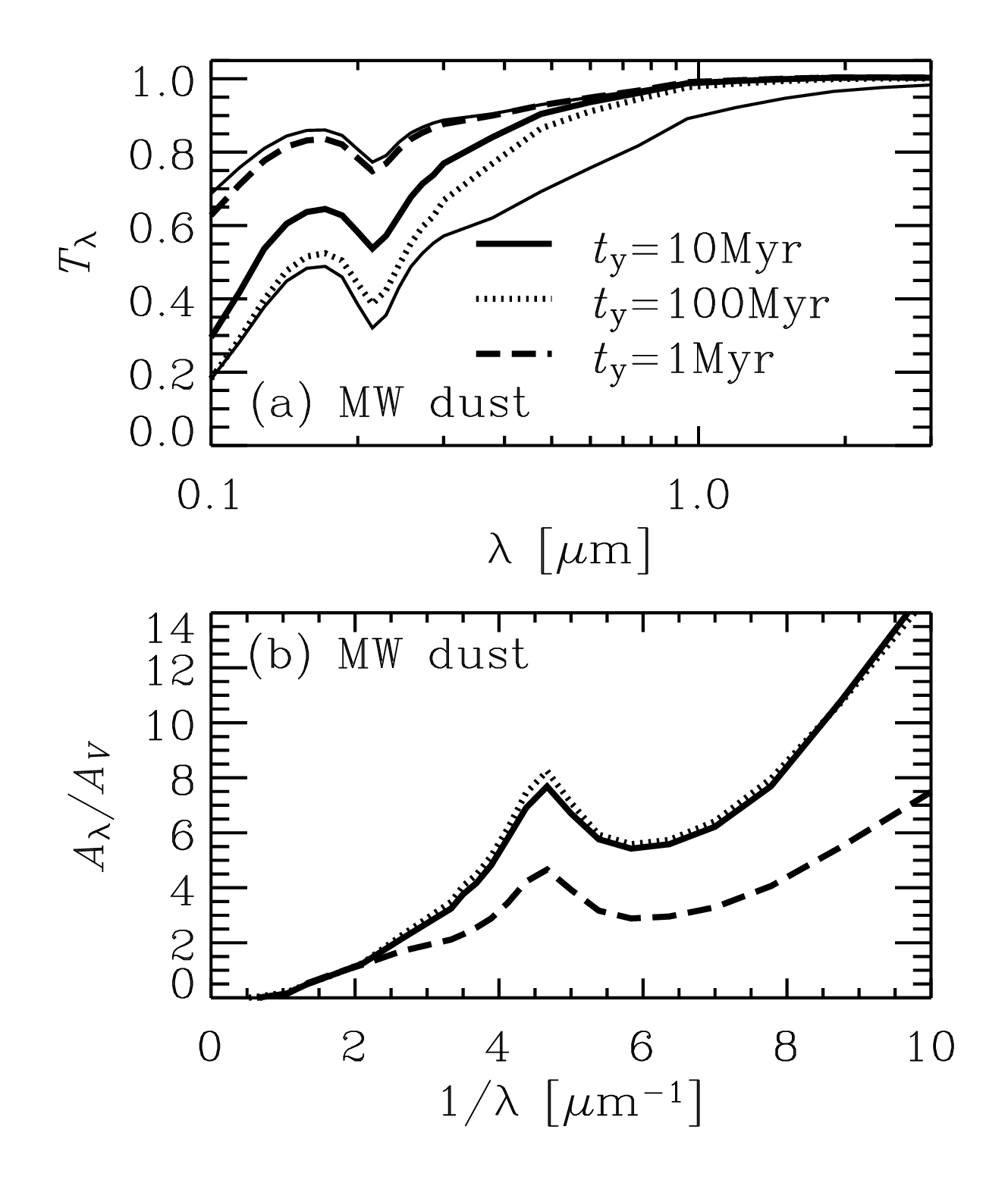}
  \caption{Dependence on the criterion of the young stellar age ($t_{\rm y}$):
    (a) transmission rates and (b) attenuation laws normalized at visual
    band. In both panels, thick solid, dotted, and dashed curves are the
    composite transmission rates/attenuation laws with $t_{\rm y}=10$ Myr, 100
    Myr, and 1 Myr, respectively. Thin solid curves in panel (a) are 
    transmission rates $T_0$ (upper) and $T_1$ (lower). 
    The Milky Way type dust is assumed.}
\end{figure}

First we change the criterion of young stars, $t_{\rm y}$. As shown in
figure~2, it affects on the luminosity fraction of young stars, $f_{\rm y}$. 
Figure~5 shows the effect of $t_{\rm y}$ on transmission rates and attenuation
laws. For the thick solid, dotted, and dashed curves, $t_{\rm y} = 10$ Myr,
100 Myr, and 1 Myr are assumed, respectively. Two thin solid curves in the
panel (a) are transmission rates $T_0$ (upper) and $T_1$ (lower). 
As $t_{\rm y}$ increases, $f_{\rm y}$ increases and approaches unity, so that
the composite transmission rate decreases and approaches $T_1$. 
On the other hand, the attenuation laws normalized at visual band for
$t_{\rm y}\ga10$ Myr are very similar. The conclusion of a steep attenuation
law is not much affected by $t_{\rm y}$ if $t_{\rm y}\ga10$ Myr.
Even for the case of $t_{\rm y} = 1$ Myr, the attenuation law is steeper than
the Calzetti's law.

\subsubsection{Layering parameter for diffuse stellar disc}

\begin{figure}
  \includegraphics[width=0.9\linewidth]{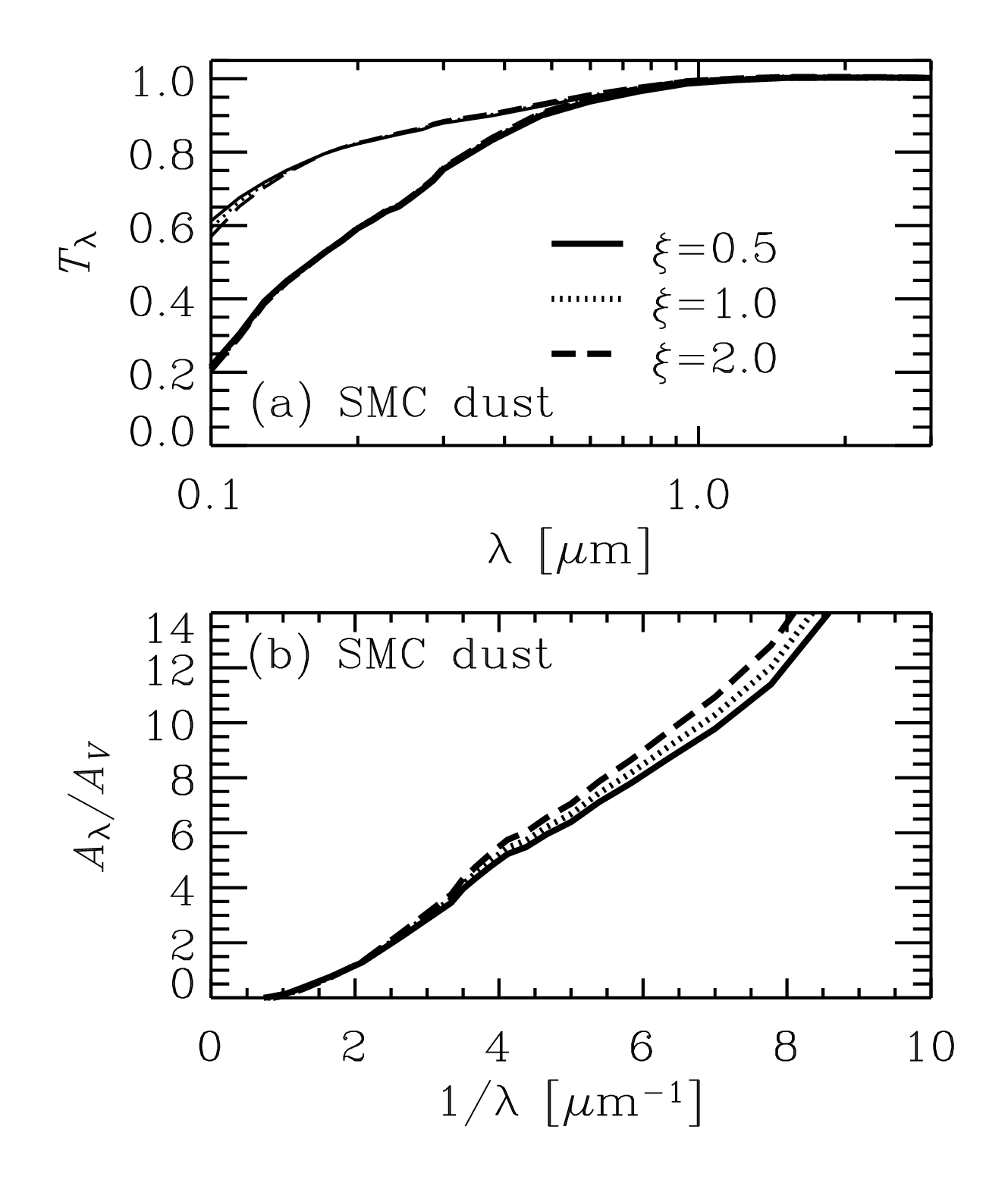}
  \caption{Dependence on the layering parameter for the diffuse stellar disc
    ($\xi$): (a) transmission rates and (b) attenuation laws normalized at
    visual band. In both panels, thick solid, dotted, and dashed curves are
    the composite transmission rates/attenuation laws with $\xi=0.5$, 1.0, and
    2.0, respectively. In panel (a), thin solid, dotted, and dashed curves are
    transmission rates $T_0$. The transmission rate $T_1$ is not affected by
    $\xi$. The Small Magellanic Cloud type dust is assumed.}
\end{figure}

Even if we change the layering parameter for diffuse stellar disc $\xi$, 
the resultant transmission rates and attenuation laws are almost not
changed as shown in figure~6 where three cases of $\xi = 0.5$ (solid), 1.0
(dotted), and 2.0 (dashed) are displayed. This is because the typical disc
considered here is not so opaque (the visual optical depth is about 0.5 in 
the uniform screen geometry). If we considered an opaque disc, the
transmission rate (equation 21) would be determined by the second term which
depends on the layering parameter $\xi$ more strongly, so that some
differences would appear.

\subsubsection{Inclination angle}

\begin{figure}
  \includegraphics[width=0.9\linewidth]{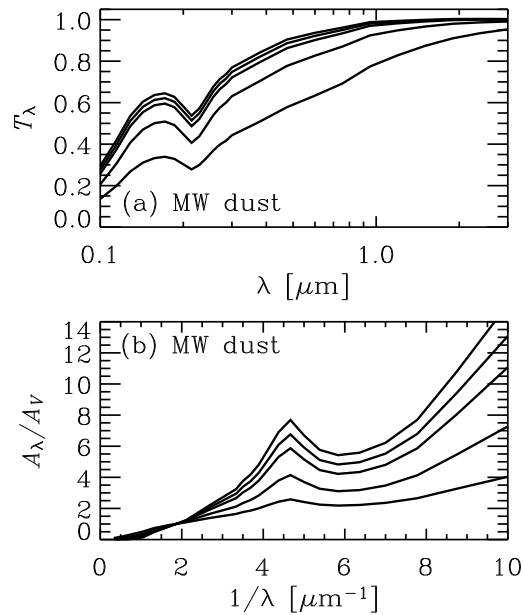}
  \caption{Dependence on the inclination angle: (a) composite transmission
    rates and (b) composite attenuation laws normalized at visual band. The
    inclination angle is $0\degr$, $29.4\degr$, $42.1\degr$, $61.1\degr$, and
    $76.9\degr$ from top to bottom in both panels. The Milky Way type dust is
    assumed.}
\end{figure}

Figure~7 shows the composite transmission rates and attenuation laws for
several cases of the inclination angle. As the inclination angle increases 
($\mu$ decreases), the disc optical depth increases because the disc apparent
thickness has a dependence of $1/\mu$. The effect of the disc edge appears
from an inclination angle larger than $89\degr$ if the ratio of the disc
radius to the vertical thickness is 100. However, the radial density structure
may affect on a highly inclined case. When we restrict ourselves to a
inclination angle of $45\degr$, for example, the difference from the face-on
case is still small, i.e. a steep attenuation law.

\subsubsection{Mean gas density}

\begin{figure}
  \includegraphics[width=0.9\linewidth]{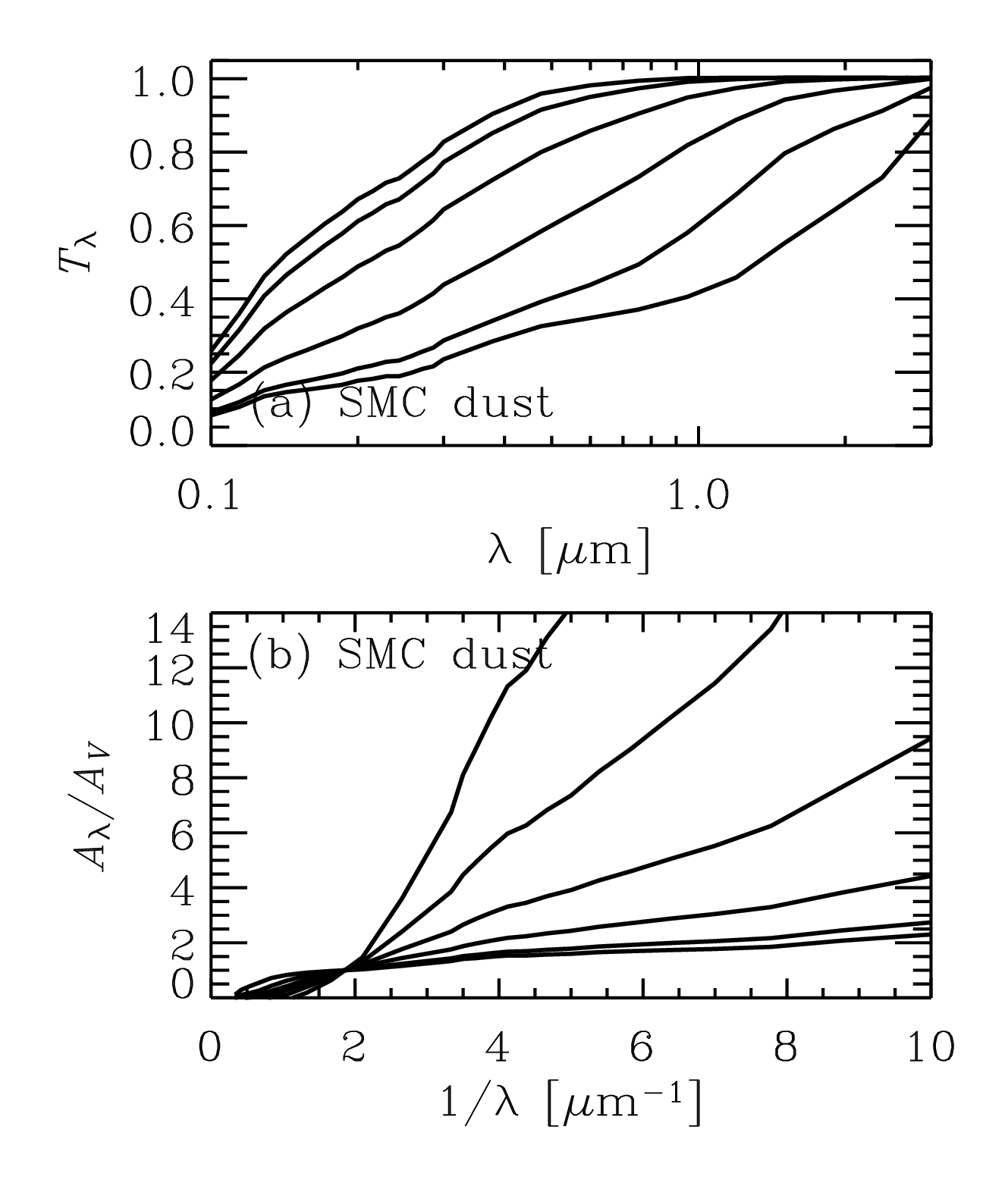}
  \caption{Dependence on the mean gas density: (a) composite transmission
    rates and (b) composite attenuation laws normalized at visual band. The
    hydrogen number density is 0.32, 0.82, 2.1, 5.5, 14.4, and 37.3 cm$^{-3}$
    from top to bottom in both panels. The Small Magellanic Cloud type dust is
    assumed.}
\end{figure}

Figure~8 shows the effect of the mean gas density on the composite
transmission rates and attenuation laws. To keep the two-phase medium
described in section 2, the hydrogen number density is limited in the range
of $n_{\rm H,wnm} \leq n_{\rm H} \leq n_{\rm H,cnm}$.
As the hydrogen number density increases, the transmission rates decreases and
the attenuation laws becomes grayer because the disc optical depth
increases.\footnote{As increasing the disc optical depth, the transmission 
rate gradually saturates at $I_*$ in equation (21). If the layering parameter 
does not depend on the wavelength, the saturation value does not,
either. Thus, we have a featureless and gray attenuation law.}
For normal disc galaxies, a mean hydrogen number density is
typically around 1 cm$^{-3}$ \citep[e.g.,][]{spi78} in which we expect a steep
attenuation law. On the other hand, a larger mean density is expected for
actively star-forming galaxies, e.g., UV bright starburst galaxies observed by
\cite{cal94}. This means that the Calzetti law can be reproduced by a high
density condition. Indeed, this is found in section 5.2.

\subsubsection{Equilibrium thermal pressure}

\begin{figure}
  \includegraphics[width=0.9\linewidth]{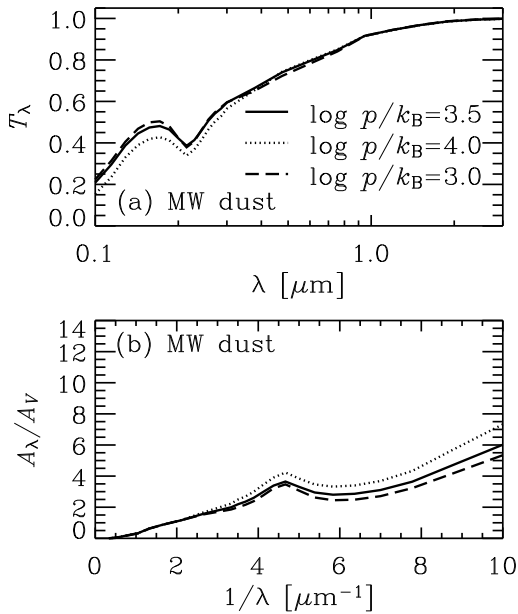}
  \caption{Dependence on the equilibrium thermal pressure ($p/k_{\rm B}$): (a)
    composite transmission rates and (b) composite attenuation laws normalized
    at visual band. In both panels, solid, dotted, and dashed curves are the
    cases with $p/k_{\rm B}=10^{3.5}$ K cm$^{-3}$, $10^{4.0}$ K cm$^{-3}$, 
    and $10^{3.0}$ K cm$^{-3}$, respectively. The Milky Way type
    dust and the mean hydrogen density of 3 cm$^{-3}$ are assumed.}
\end{figure}

When we change the equilibrium thermal pressure, we should direct our
attention to the local minimum and maximum pressures. Since we stand on a
two-phase ISM model described in section 2, we restrict ourselves within
the range between $p/k_{\rm B}=10^{3}$ K cm$^{-3}$ and $10^{4}$ K cm$^{-3}$
expected that a two-phase medium is established in the environment of the
Milky Way's ISM \citep[figure~1, see also][]{wol03}. We show the dependence on
the thermal pressure in figure~9 where the mean hydrogen density of 3
cm$^{-3}$ is assumed in order to have a clump filling fraction larger than
zero for the $p/k_{\rm B}=10^{3}$ K cm$^{-3}$ case. If we adopted the mean
density of 1 cm$^{-3}$, we would have $n_{\rm H} = n_{\rm H,wnm}$ (see eq.1),
and then, $f_{\rm cl}=0$ for the $p/k_{\rm B}=10^{3}$ K cm$^{-3}$ case. 
The pressure affects on the attenuation law by two ways: the density contrast
between clumps and the inter-clump medium, and the clump optical depth through
the size (i.e.\ Jeans length). Approximately, the density contrast is
proportional to $p^{0.4}$ found from equations (1) and (2), and the clump
optical depth is proportional to $p^{0.5}$ found from equation (3). 
Thus, the effect of the pressure is not so large as shown in figure~9 for the
limited range of the value considered here.

\subsubsection{Dust-to-gas mass ratio}

\begin{figure}
  \includegraphics[width=0.9\linewidth]{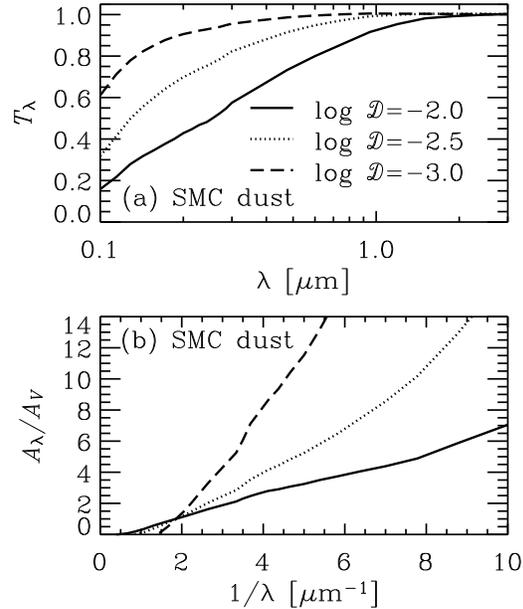}
  \caption{Dependence on the dust-to-gas mass ratio ($\cal D$): (a) composite 
    transmission rates and (b) composite attenuation laws normalized at visual
    band. In both panels, solid, dotted, and dashed curves are the cases with 
    ${\cal D} = 10^{-2.0}$, $10^{-2.5}$, and $10^{-3.0}$, respectively. The
    Small Magellanic Cloud type dust and the mean hydrogen density of 3
    cm$^{-3}$ are assumed.}
\end{figure}

Figure~10 shows the effect of the dust-to-gas mass ratio. Here we adopt the
mean hydrogen density of 3 cm$^{-3}$ to avoid having a too small disc optical
depth for the case of ${\cal D}=10^{-3}$, otherwise we have almost zero
attenuation at visual band which is the normalization of the presentation.
We find that a steeper attenuation law as ${\cal D}$ decreases because the
disc optical thickness decreases. Conversely, we can make a grayer attenuation
law if ${\cal D}$ increases. However, it is unlikely that many galaxies have a 
much higher dust-to-gas ratio than that of the Milky Way.

\subsubsection{Global geometry}

\begin{figure}
  \includegraphics[width=0.9\linewidth]{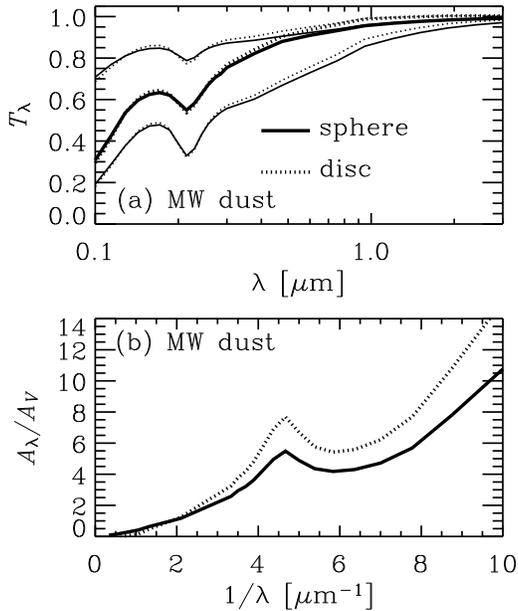}
  \caption{Comparison with a disc geometry and a spherical geometry: (a)
    transmission rates and (b) attenuation laws normalized at visual band. The
    solid curves are the spherical case and the dotted curves are the disc
    case. The thin curves in the panel (a) are transmission rates $T_0$
    (upper) and $T_1$ (lower). The thick curves in both panels are the
    composite transmission rates/attenuation laws. 
    The Milky Way type dust is assumed.}
\end{figure}

We examine the effect of the global geometry. Suppose a sphere having a
two-phase medium with an equilibrium thermal pressure and a mean gas
density. Young stars are confined in clumps, and clumps and older stars
distribute uniformly in the sphere.
When we consider isotropic scatterings and assume that the
scattered photons are also distributed uniformly in the sphere, the
transmission rate for diffuse sources can be analytically expressed as
equation (10) with the effective optical depth of $\kappa_{\rm eff}R$ and the
effective albedo $\omega_{\rm eff}$. The radius of the sphere $R$ is set to be
equal to the dusty disc $h_{\rm d}$ for the comparison. The clump optical
depth and albedo are also given by equations (3) and (9). Therefore, we have
the transmission rate of the clumpy sphere as 
\begin{equation}
  T^{\rm sph} = f_{\rm y} T_1^{\rm sph} + (1-f_{\rm y}) T_0^{\rm sph}\,,
\end{equation}
where the transmission rate for the diffuse source is  
\begin{equation}
  T_0^{\rm sph} = P_{\rm esc}(\kappa_{\rm eff}h_{\rm d},\omega_{\rm eff})\,,
\end{equation}
and the transmission rate for the clumpy source is 
\begin{equation}
  T_1^{\rm sph} = P_{\rm esc}(\tau_{\rm cl},\omega_{\rm cl})
  P_{\rm esc}(\kappa_{\rm eff}h_{\rm d},\omega_{\rm eff})\,.
\end{equation}

Figure~11 shows the comparison of the disc and spherical geometries. We have
assumed the standard set tabulated in table~1 in this comparison. For the disc
geometry, we adopted $\xi=0.5$ since this choice is not so important as shown
in section 5.1.2. We find very similar transmission rates. Differences
appear in longer wavelengths where the optical depth is small. This
differences are caused by the scattering. In the face-on disc geometry, the path
length of the photons scattered into the observer's (face-on) line of sight is
shorter than the original length. On the other hand, the scattered path length
is longer than the original one in the spherical geometry. Thus, the face-on
disc is more transparent for the scattered photons than the spherical
geometry. This leads to the difference of the attenuation law normalized at
visual band because of the different normalization.

\subsubsection{Grayer dust in clumps}

\begin{figure}
  \includegraphics[width=0.9\linewidth]{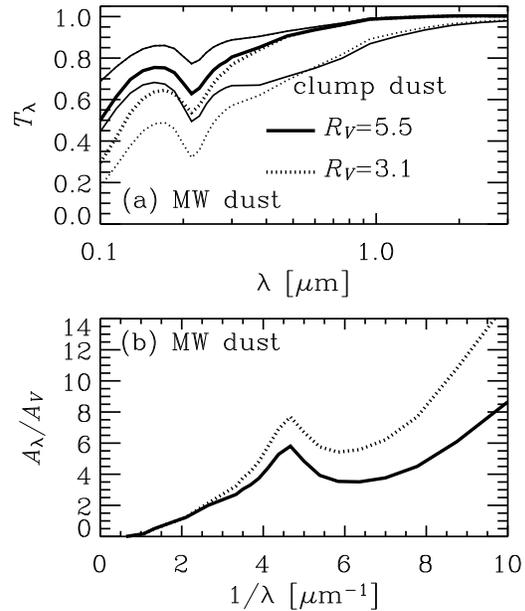}
  \caption{Different dust property in clumps: (a) transmission rates and (b)
    attenuation laws normalized at visual band. The solid curves are the case
    that the extinction law of $R_V=5.5$ is assumed only for dust in clumps,
    and the dotted curves are the case that the clump dust is the same as the
    diffuse dust ($R_V=3.1$). The thin curves in the panel (a) are
    transmission rates $T_0$ (upper) and $T_1$ (lower). The thick curves in
    both panels are the composite transmission rates/attenuation laws. The
    Milky Way type dust is assumed.}
\end{figure}

The observed extinction laws in the Milky Way and Magellanic Clouds show a
very large variety among lines of sight and are expected to depend on the
environment \citep[e.g.,][]{fit99,gor03}. Especially, the dust property in
star-forming regions (i.e. clumps) may be different from that in diffuse
medium. Indeed, \cite{car89} showed that the extinction laws along some lines
of sight toward dense clouds have a large value of $R_V\equiv A_V/E(B-V)$; a
grayer extinction law. For example, they found $R_V=5.5$ toward the Orion
nebula, in contrast, $R_V=3.1$ for the average extinction law. 
Such a large $R_V$ was attributed to the grain growth in dense medium. 
On the other hand, \cite{rac02} reported that there is no correlation between
$R_V$ and the molecular fraction along translucent lines of sight.
Although a systematic difference of the dust property in dense medium (a
grayer extinction law) is not completely established, the possible effect is
worth examining.

We assume that the dust in clumps is grayer than the diffuse dust; the
extinction cross section ($k_{\rm d,\lambda}$) in clumps is set to be the
extinction law with $R_V=5.5$ by Cardelli et al.~(1989). The extinction cross
section for the inter-clump dust is the same as introduced in section
3.1. This corresponds to the case of $R_V=3.1$. The scattering albedo
($\omega_{\rm d,\lambda}$) and the asymmetry parameter ($g_{\rm d,\lambda}$)
are not changed because we do not have enough informations about them for the
case of $R_V=5.5$. If the dust size of $R_V=5.5$ is larger than that of
$R_V=3.1$, the adopted scattering albedo for $R_V=5.5$ may be smaller than the
real values in optical wavelengths. It leads a underestimation of the
transmission rate $T_1$. However, the composite attenuation at the visual band
is dominated by $T_0$ which is determined by the $R_V=3.1$ dust. 
Thus, the effect is not important.

Figure~12 shows the case assuming the grayer dust of $R_V=5.5$ in clumps as
solid curves. For the dotted curves, the clump dust is the same as the diffuse
dust which has always $R_V=3.1$. From the top panel, we find that the
transmission rate $T_1$ of $R_V=5.5$ becomes larger and grayer than that of
$R_V=3.1$. This makes the composite attenuation law grayer (bottom panel). 
However, it is still steeper than the extinction law (of $R_V=3.1$) and the
Calzetti law.

\subsection{Calzetti law}

\begin{figure*}
  \includegraphics[width=0.9\linewidth]{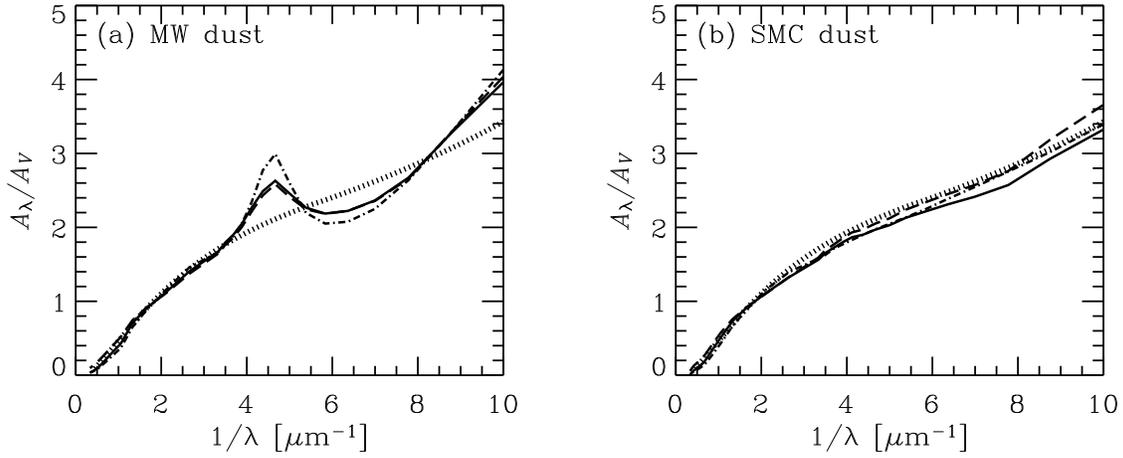}
  \caption{Comparison of the attenuation laws with the Calzetti law: (a) Milky
    Way type dust and (b) Small Magellanic Cloud type dust. The dotted curve
    is the Calzetti law. The solid curves are the composite attenuation laws
    for the face-on case with the mean hydrogen number densities of (a) 5.5
    cm$^{-3}$ and (b) 8.9 cm$^{-3}$. The dashed curves are the composite
    attenuation laws of the mean hydrogen density of 1 cm$^{-3}$ for the
    inclination angle of (a) $76.9\degr$ and (b) $80.7\degr$. The dot-dashed
    curves are the face-on attenuation law for $T_1$ with the mean hydrogen
    densities of (a) 3.4 cm$^{-3}$ and (b) 8.9 cm$^{-3}$.}
\end{figure*}

The attenuation law obtained by \cite{cal94} from observations of the UV
bright starburst galaxies is grayer than the extinction laws of the Milky
Way and the Magellanic Clouds. Here we try to reproduce the Calzetti law in
the framework of this paper: the attenuation law through the two-phase ISM and
the age-selective attenuation.

Although we have obtained a steep attenuation law for normal disc galaxies in
section 4, we can also make a grayer attenuation law as seen in section 5.1,
for example, a larger mean density. Figure~13 shows a good agreement between
the Calzetti law (dotted) with our composite attenuation law for the face-on
case with a mean hydrogen density of 5--10 cm$^{-3}$ (solid). Active
star-forming galaxies are likely to have a large amount of gas, and the gas
density should exceed a certain critical value \citep{ken89}. The hydrogen
density of 5--10 cm$^{-3}$ corresponds to 50--100 $M_\odot {\rm pc}^{-2}$ for
the disc half height of 150 pc. Indeed, this gas column density greatly
exceeds the critical density for the star formation found by \cite{ken89}.

There are two alternative ways to reproduce the Calzetti law by our model: a
highly inclined disc and a system dominated by young stars. 
Even in the case of the mean hydrogen density of 1 cm$^{-3}$, the composite
attenuation laws with an inclination angle of 75--80$\degr$ (dashed curves in
figure~13) also show a good agreement with the Calzetti law. However, this is
unlikely because there is no systematic bias toward very high inclinations in
the sample galaxies of \cite{cal94}. Since the galaxies really show an active
star formation, the luminosity may be dominated by young stars. In this case,
the attenuation law with $f_{\rm y}=1$ (i.e. attenuation law produced only by
$T_1$) is suitable. The dot-dashed curves show such cases with a mean hydrogen
density of 3--10 cm$^{-3}$ (face-on). Again, a larger density is preferred.

\subsection{Charlot and Fall prescription}

\begin{figure*}
  \includegraphics[width=0.9\linewidth]{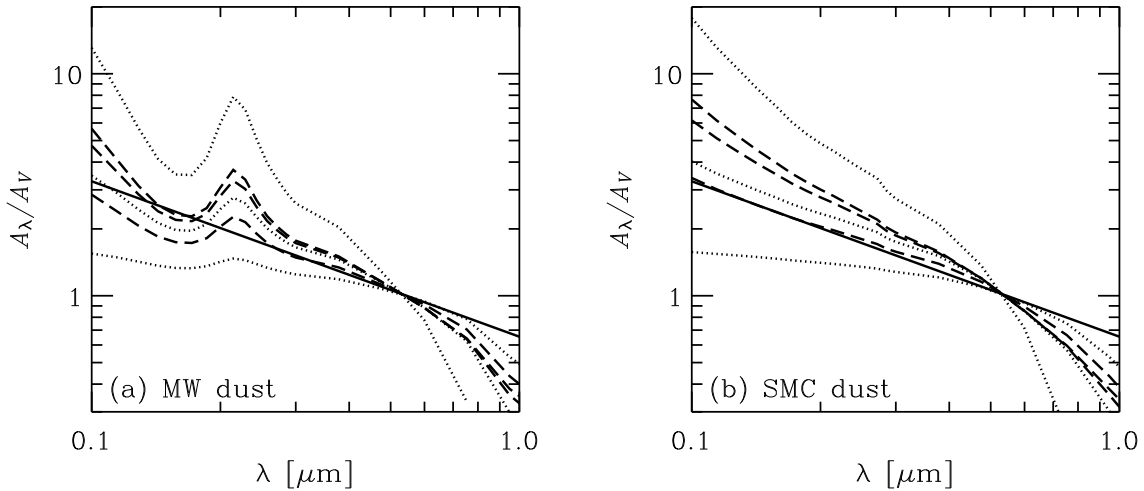}
  \caption{Comparison of attenuation laws with the prescription by
    Charlot \& Fall (2000): (a) Milky Way type dust and (b) Small Magellanic
    Cloud type dust. The solid line is the power-law attenuation law by
    Chalrot \& Fall (2000); $A_\lambda \propto \lambda^{-0.7}$. The dotted and
    dashed curves are the attenuation laws corresponding transmission rates
    $T_0$ and $T_1$, respectively. The mean hydrogen number densities are
    0.51, 2.1, and 8.9 cm$^{-3}$ from top to bottom for both types of curves
    and in both panels.}
\end{figure*}

\cite{cha00} proposed a simple prescription of the attenuation law with the
age-selective attenuation and explained a lot of observational properties of
the UV bright starburst galaxies. They adopted the age-selective attenuation by
a similar way to us: the optical depth for old stars is only that by dust
diffusely distributed in the ISM $\tau_{\rm ISM}$ and the optical depth for
young stars is the sum of $\tau_{\rm ISM}$ and the optical depth by dust in
the birth cloud $\tau_{\rm BC}$. Then, they simply assumed that the wavelength
dependences of $\tau_{\rm ISM}$ and $\tau_{\rm BC}$ are the same: 
$\lambda^{-0.7}$. We examine this point.

In figure~14, we show the normalized attenuation laws for $T_1$ (dashed) and
$T_0$ (dotted) with several mean hydrogen densities, and compare them with the
power-law attenuation with the index -0.7 (solid) proposed by \cite{cha00}. 
We find that (1) the slopes of the attenuation laws for the embedded young
stars ($T_1$) and for the diffuse stars ($T_0$) are not the same. (2) The
slopes decreases, especially for the diffuse stars, as the mean gas
density increases. (3) The slopes for both stellar populations are similar to
each other in a case with a mean hydrogen density of $\sim 5$ cm$^{-3}$, in
which the composite attenuation law is also similar to the Calzetti law. 
Therefore, the Charlot and Fall prescription and also the Calzetti law can be
interpreted as a special case of a larger density (i.e. starburst) disc in
the framework presented here.

\section{Conclusion}

The attenuation laws through the interstellar medium (ISM) with clumpy
distributions of stars and dust was investigated in this paper. As an
application of the multi-phase ISM model, we introduced a physical model of
the clumpiness of the medium (i.e.\ dust distribution). In the model, the two
parameters determining the clumpiness, the density contrast between clumps and
the inter-clump medium and the volume filling fraction of clumps, were
translated into two physical quantities in the ISM: the equilibrium thermal
pressure and the mean gas density. 
The radiative transfer through the clumpy ISM was treated in a 1-D
plain-parallel geometry by the mega-grain approximation developed by
\cite{var99}. For the clumpiness of the stellar distribution, we considered
two stellar populations: older stars diffusely and smoothly distributed in the 
medium and younger stars embedded in clumps (i.e.\ birth clouds). This provides
the age-selective attenuation \citep{sil98}; young stars are surrounded by a
denser dusty medium and suffer a stronger attenuation. The transmission rate of
the entire radiation from the two stellar populations is a composite of the
transmission rates of each individual population with a luminosity
weight. The stellar population dominating the radiation changes from older
stars to younger stars as the wavelength decreases. With this fact, the
age-selective attenuation makes the attenuation law steep; the composite
attenuation rapidly increases from small attenuation for older stars at a
long wavelength to large attenuation for younger stars at a short
wavelength. Therefore, the attenuation law expected for normal disc galaxies
is much steeper than the Calzetti law \citep{cal94} and the extinction laws in
the Milky Way and the Magellanic Clouds. 
After an extensive survey of the parameter space, it was
shown that the steep attenuation law is very robust for the parameter set
expected in normal disc galaxies. Furthermore, the Calzetti law was
reproduced in the framework presented here with a larger gas density than that
in normal disc galaxies. Such a large gas density is very likely for the
starburst galaxies. Finally, we examined the validity of a simple prescription
of the attenuation law proposed by \cite{cha00}. Then, we found that their
prescription (and also the Calzetti law) is a special case with a larger
density disc in our model.

Interestingly, the UV color (or spectral slope) of normal disc galaxies is
systematically redder than that of UV bright starbursts at a fixed flux ratio
of the UV to the infrared (or fixed UV attenuation)
\citep[e.g.,][]{bel02}. A first result of GALEX also shows such trend
\citep{bua04}. One of the explanations of the trend is that normal disc
galaxies have intrinsically red colors because of their quiescent star-forming
activity \citep{kon04}. However, a steep attenuation law presented here can
make galaxies red with small attenuation, and then, can be an alternative
interpretation of the trend. Nowadays, GALEX is providing us with the UV data,
including spectra, of a large number of normal disc galaxies. In the near
future, GALEX will conclude whether the attenuation law of normal disk
galaxies is steep or not.

\section*{Acknowledgments}
The author thanks Taishi Nakamoto, Tsutomu T.\ Takeuchi, Pasquale Panuzzo,
Veronique Buat, Denis Burgarella, Jean-Michel Deharveng, Hideyuki Kamaya,
Toshinobu Takagi, and Nobuo Arimoto for discussions and encouragements. 
The author also thanks the anonymous referee for useful suggestions. 
The author is supported by the JSPS Postdoctoral Fellowships for Research
Abroad.

\label{lastpage}

\begin{thebibliography}{}
\bibitem[Baes \& Dejonghe(2001a)]{bae01a}
Baes, M., Dejonghe, H., 2001a, MNRAS, 326, 722

\bibitem[Baes \& Dejonghe(2001b)]{bae01b}
Baes, M., Dejonghe, H., 2001b, MNRAS, 326, 733

\bibitem[Bell(2002)]{bel02}
Bell, E. F., 2002, ApJ, 577, 150

\bibitem[Bianchi, et al.(2000)]{bia00}
Bianchi, S., Ferrara, A., Davies, J. I., Alton, P. B., 2000, MNRAS, 311, 601

\bibitem[Binney \& Merrifield(1998)]{bin98}
Binney, J., Merrifield, M., 1998, Galactic Astronomy, Princeton University
Press, NJ

\bibitem[Blitz \& Shu(1980)]{bli80}
Blitz, L., Shu, F. H., 1980, ApJ, 238, 148

\bibitem[Boulares \& Cox(1990)]{bou90}
Boulares, A., Cox, D. P., 1990, ApJ, 365, 544

\bibitem[Bruzual et al.(1988)]{bru88}
Bruzual, A. G., Magris, G. C., Calvet, N., 1988, ApJ, 333, 673

\bibitem[Buat et al.(2002)]{bua02}
Buat, V., Bugarella, D., Deharveng, J.-M., Kunth, D., 2002, A\&A, 393, 33

\bibitem[Buat et al.(2005)]{bua04}
Buat, V., Iglesias-P{\'a}ramo, J., Seibert, M., Burgarella, D., et al., 
2005, ApJ, 619, L51

\bibitem[Calzetti(2001)]{cal01}
Calzetti, D., 2001, PASP, 113, 1449

\bibitem[Calzetti et al.(1994)]{cal94}
Calzetti, D., Kinney, A. L., Storchi-Bergmann, T., 1994, ApJ, 429, 582

\bibitem[Cardelli, Clayton, \& Mathis(1989)]{car89}
Cardelli, J. A., Clayton, G. C., Mathis, J. S., 1989, ApJ, 345, 245

\bibitem[Charlot \& Fall(2000)]{cha00}
Charlot, S., Fall, S. M., 2000, ApJ, 539, 718

\bibitem[de Bruijne(1999)]{deb99}
de Bruijne, J. H. J., 1999, MNRAS, 310, 585

\bibitem[Di Bartolomeo et al.(1995)]{dib95}
Di Bartolomeo, A., Barbaro, G., Perinotto, M., 1995, MNRAS, 277, 1279

\bibitem[Draine(2003)]{dra03}
Draine, B. T., 2003, ApJ, 598, 1017

\bibitem[Ferrara et al.(1999)]{fer99}
Ferrara, A., Bianchi, S., Cimatti, A., Giovanardi, C., 1999, ApJS, 123, 437

\bibitem[Field et al.(1969)]{fie69}
Field, G. B., Goldsmith, D. W., Habing, H. J., 1969, ApJ, 155, L149

\bibitem[Fischera et al.(2003)]{fis03}
Fischera, J., Dopita, M. A., Sutherland, R. S., 2003, ApJ, 599, L21

\bibitem[Fischera \& Dopita(2005)]{fis04}
Fischera, J., Dopita, M. A., 2005, ApJ, 619, 340

\bibitem[Fitzpatrick(1999)]{fit99}
Fitzpatrick, E. L., 1999, PASP, 111, 63

\bibitem[Gordon et al.(1997)]{gor97}
Gordon, K. D., Calzetti, D., Witt, A. N., 1997, ApJ, 487, 625

\bibitem[Gordon et al.(2000)]{gor00}
Gordon, K. D., Clayton, G. C., Witt, A. N., Misselt, K. A., 2000, ApJ, 533,
236

\bibitem[Gordon et al.(2003)]{gor03}
Gordon, K. D., Clayton, G. C., Misselt, K. A., Landolt, A. U., Wolff, M. J.,
2003, ApJ, 594, 279

\bibitem[Granato et al.(2000)]{gra00}
Granato, G. L., Lacey, C. G., Silva, L., Bressan, A., Baugh, C. M., Cole, S.,
Frenk, C. S., 2000, ApJ, 542, 710

\bibitem[Heiles \& Troland(2003)]{hei03}
Heiles, C., Troland, T. H., 2003, ApJ, 586, 1067

\bibitem[Henyey \& Greenstein(1941)]{hen41}
Henyey, L. C., Greenstein, J. L., 1941, ApJ, 93, 70

\bibitem[Hobson \& Padman(1993)]{hob93}
Hobson, M. P., Padman, R., 1993, MNRAS, 264, 161

\bibitem[Kennicutt(1989)]{ken89}
Kennicutt, R. C., 1989, ApJ, 344, 685

\bibitem[Kong et al.(2004)]{kon04}
Kong, X., Charlot, S., Brinchmann, J., Fall, S. M., 2004, MNRAS, 349, 769 

\bibitem[Koyama \& Inutsuka(2000)]{koy00}
Koyama, H., Inutsuka, S.-I., 2000, ApJ, 532, 980

\bibitem[Larson(1981)]{lar81}
Larson, R. B., 1981, MNRAS, 194, 809

\bibitem[Leitherer et al.(2002)]{lei02}
Leitherer, C., Li, I.-H., Calzetti, D., Heckman, T. M., 2002, ApJS, 140, 303

\bibitem[Martin et al.(2005)]{mar04}
Martin, D. C., Fanson, J., Schiminovich, D., Morrissey, P., et al., 
2005, ApJ, 619, L1

\bibitem[Matthews \& Wood(2001)]{mat01}
Matthews, L. D., Wood, K., 2001, ApJ, 548, 150

\bibitem[Mckee \& Ostriker(1977)]{mck77}
McKee, C. F., Ostriker, J. P., 1977, ApJ, 218, 148

\bibitem[Mihalas \& Weibel-Mihalas(1999)]{mih99}
Mihalas, D., Weibel-Mihalas, B., 1999, Foundations of Radiation Hydrodynamics, 
Dover, New York

\bibitem[Myers(1978)]{mye78}
Myers, P. C., 1978, ApJ, 225, 380

\bibitem[Natta \& Panagia(1984)]{nat84}
Natta, A., Panagia, N., 1984, ApJ, 287, 228

\bibitem[Neufeld(1991)]{neu91}
Neufeld, D. A., 1991, ApJ, 370, L85

\bibitem[Ng(1974)]{ng74}
Ng, K.-C., 1974, J. Chem. Phys., 61, 2680

\bibitem[Olson et al.(1986)]{ols86}
Olson, G. L., Auer, L. H., Buchler, J. R., 1986, JQSRT, 35, 431

\bibitem[Pierini et al.(2004)]{pie04}
Pierini, D., Gordon, K. D., Witt, A. N., Madsen, G. J., 2004, ApJ, 617, 1022

\bibitem[Rachford et al.(2002)]{rac02}
Rachford, B. L., Snow, T. P., Tumlinson, J., Shull, J. M., et al., 2002, ApJ,
577, 221

\bibitem[Silva et al.(1998)]{sil98}
Silva, L., Granato, G. L., Bressan, A., Danese, L., 1998, ApJ, 509, 103

\bibitem[Spitzer(1978)]{spi78}
Spitzer, L., 1978, Physical Processes in the Interstellar Medium, Wiley, New
York

\bibitem[Tuffs et al.(2004)]{tuf04}
Tuffs, R. J., Popescu, C. C., V{\" o}lk, H. J., Kylafis, N. D., Dopita, M. A.,
2004, A\&A, 419, 821

\bibitem[V{\'a}rosi \& Dwek(1999)]{var99}
V{\' a}rosi, F., Dwek, E., 1999, ApJ, 523, 265

\bibitem[Witt \& Gordon(1996)]{wit96}
Witt, A. N., Gordon, K. D., 1996, ApJ, 463, 681

\bibitem[Witt \& Gordon(2000)]{wit00}
Witt, A. N., Gordon, K. D., 2000, ApJ, 528, 799

\bibitem[Wolfire et al.(1995)]{wol95}
Wolfire, M. G., Hollenbach, D., McKee, C. F., Tielens, A. G. G. M., Bakes,
E. L. O., 1995, ApJ, 443, 152

\bibitem[Wolfire et al.(2003)]{wol03}
Wolfire, M. G., McKee, C. F., Hollenbach, D., Tielens, A. G. G. M., 2003, ApJ,
587, 278

\end{thebibliography}
\end{document}